\documentclass[pra, aps, showpacs, superscriptaddress, twocolumn]{revtex4-2}
\usepackage{graphicx,amsmath,amssymb,color}
\usepackage{physics}
\usepackage{graphicx}
\usepackage{epstopdf}
\usepackage{float}
\usepackage{xcolor}
\usepackage[colorlinks, citecolor={blue!50!black}, urlcolor={blue!50!black}, linkcolor={red!50!black}]{hyperref}

\begin{document}

\title{Multiphoton blockade and antibunching in an optical cavity coupled with dipole-dipole interacting $\Lambda$-type atoms}

\author{Zeshan Haider}
\email[Present Adress: National Institute of Lasers and Optronics, PIEAS, Nilore, Islamabad  $45650$, Pakistan.]{}
\affiliation{Department of Physics and Applied Mathematics, Pakistan Institute of Engineering and Applied Sciences (PIEAS), Nilore, Islamabad  $45650$, Pakistan.}
\author{Shahid Qamar}
\affiliation{Department of Physics and Applied Mathematics, Pakistan Institute of Engineering and Applied Sciences (PIEAS), Nilore, Islamabad  $45650$, Pakistan.}
\affiliation{Center for Mathematical Sciences, PIEAS, Nilore, Islamabad $45650$, Pakistan.}
\author{Muhammad Irfan}
\affiliation{Department of Physics and Applied Mathematics, Pakistan Institute of Engineering and Applied Sciences (PIEAS), Nilore, Islamabad  $45650$, Pakistan.}
\affiliation{Center for Mathematical Sciences, PIEAS, Nilore, Islamabad $45650$, Pakistan.}

\date{\today}

\begin{abstract}
We study multiphoton blockade effects in a single-mode cavity interacting with two three-level atoms in $\Lambda$-configuration having position-dependent atom-field coupling.
We consider the effects of dipole-dipole interaction (DDI) between the three-level atoms and show how the presence of DDI strongly influences the multiphoton blockade.
For symmetric coupling of the atoms with the field, the DDI induces an asymmetry in the emission spectra as a function of pump field detuning.
At positive detuning, the single-photon blockade gets stronger as a function of DDI strength, leading to photon antibunching.
However, it becomes weaker at negative detuning and can also completely vanish.
We show that this vanishing single-photon blockade is associated with a strong two-photon blockade, leading to two-photon bunching.
Therefore, by just tuning the frequency of the pump field, we can achieve two very distinct features.
We also study the effects of DDI when the atoms are asymmetrically coupled with the field and show that the proposed system exhibits two-photon bunching.
We believe our results are important for the experimental realization of such systems where DDI may be present.
\end{abstract}

\maketitle
\section{introduction}
Photon antibunching is a purely quantum mechanical effect that has no classical counterpart~\cite{paul_1982,scully}.
In photon antibunching, a stream of temporally well-spaced single photons can be generated by a blockade of two or more photon generation, an effect known as single-photon blockade.
Similarly, it is also possible to generate the non-classical photon pairs by blockade of a third photon generation (two-photon blockade).
The phenomenon of the multiphoton blockade arises due to the anharmonicity of the low-laying collective states of the system (dressed states). 
These non-classical states of light have many applications in quantum communication~\cite{Gisin2007}, quantum metrology~\cite{Giovannetti2011}, quantum computing~\cite{OBrien2007}, among others.

Photon blockade effect has been studied extensively in various physical systems including cavity-QED~\cite{Imamo1997, SHAMAILOV2010766, Hamsen2017, Hamsen2018, Zigeng2021, Ming2022, Kowalewska2019}, circuit-QED~\cite{Houck2007, Lang2011, Hoffman2011, Liu2014, fink2017, Vaneph2018}, trapped atoms~\cite{Birnbaum2005}, quantum dots~\cite{Faraon2008, Snijders2018}, optomechanical systems~\cite{Rabl2011, Wang_2020},  magnomechanical systems~\cite{Zhao2020}, diamond nanophotonic cavity~\cite{Knall2022}, quantum wells~\cite{Kyriienko2014}, semiconducting transition metal dichalcogenides~\cite{Kyriienko2020, Emmanuele2020}, among others.
An atom coupled with a cavity is an ideal system to realize multiphoton blockade~\cite{SHAMAILOV2010766, wudeng_2015, Hamsen2017, Hamsen2018, MAVROGORDATOS2021126791} because of the strong atom-field coupling.
Many interesting studies have been carried out in recent years studying photon blockade in two-level and three-level atoms coupled with single-mode cavities~\cite{Hamsen2017, Tang2019, Han2018, Guo2022}.
Furthermore, the single two-level atom case was extended to the two-photon Jaynes-Cummings model, showing an enhanced photon blockade~\cite{Zou2020}.
Moreover, it is also shown that multi-atom cavity-QED has very interesting effects on photon statistics~\cite{Greentree2000, Lin2015, Bajcsy2013, Chen2022}. 

Recently, Zhu {\it et al.} considered coherently driven two two-level atoms with position-dependent coupling in a single-mode cavity~\cite{Zhu2017}.
They showed that single and two-photon blockades can be observed simultaneously, under appropriate conditions where the location of atoms plays an important role.
In the same system, Pleinert {\it et al.} have shown that a strong atom-field coupling regime leads to correlated emission surpassing the superradiant emission, a phenomenon they termed hyperradiance~\cite{Pleinert2017}.
Radulaski {\it et al.} also independently considered a similar system in the bad cavity limit and showed three different mechanics of photon blockade~\cite{Radulaski2017}.
Inspired by these results, a number of subsequent interesting studies were carried out~\cite{Lin2019, Han2020, Li2021, Zhu2021, Xia2022, Huang2021, Zhang_2022}.
For instance, it is shown that replacing two-level atoms with three-level atoms in a cascade configuration enhances the strength of two-photon blockade~\cite{Lin2019, Han2020, Li2021}.
This is because of the inherent interesting phenomenon of electromagnetic-induced transparency in a three-level atomic system.

It is well known that depending upon the separation between the atoms, the dipole-dipole interaction (DDI) has important consequences on the energy spectrum and photon blockade~\cite{Qu2020, Devi2020, Zheng2016}.
It is therefore important to investigate the effects of DDI in photon blockade studies in such schemes.
Zhu {\it et al.} included DDI between two two-level atoms coupled with a single-mode field~\cite{Zhu2021}.
The presence of DDI induces a shift in energies of the dressed states, resulting in improved photon number and correlation functions as compared to the case where DDI is absent~\cite{Zhu2017}.
From Ref.~\cite{Lin2019}, we know that a three-level atomic system enhances photon blockade, while the presence of DDI in two-level atomic systems results in improved photon number and correlation function values due to the shift in energies of the dressed states~\cite{Zhu2021}.
It is, therefore, important to investigate DDI in two three-level atoms coupled with a single-mode cavity.
To the best of our knowledge, such a study is not reported so far.
In this paper, we consider two three-level atoms in $\Lambda-$configuration having DDI and coupled to a single-mode cavity.
The atom-field coupling is assumed to be position-dependent.
We show that the presence of DDI plays an important role.
For instance, for symmetric coupling of two atoms, it induces an asymmetry in the emission peaks in the detuning space improving single-photon blockade at one detuning while suppressing at the other.
At positive pump field detuning, the mean photon number stays almost constant for increasing DDI strength while the second-order correlation function gets weaker, leading to a stronger single-photon blockade.
At negative detuning, the mean photon number decreases with increasing DDI strength accompanied by a stronger second-order correlation function, resulting in a weaker single-photon blockade.
We show that for a proper choice of parameters the single-photon blockade vanishes, and we obtain a strong two-photon blockade, showing the two-photon-bunching.
Therefore, the proposed system promises the realization of a single-photon source at positive detuning and a two-photon source at negative detuning which can be controlled by tuning the frequency of the pump field.
Finally, we discuss the effects of DDI on correlation functions for asymmetric coupling of the atoms with the field.
In this regime, our system exhibits two-photon bunching for a proper combination of drive field and DDI strength. 
\begin{figure}[htb]
	\includegraphics[scale=0.5]{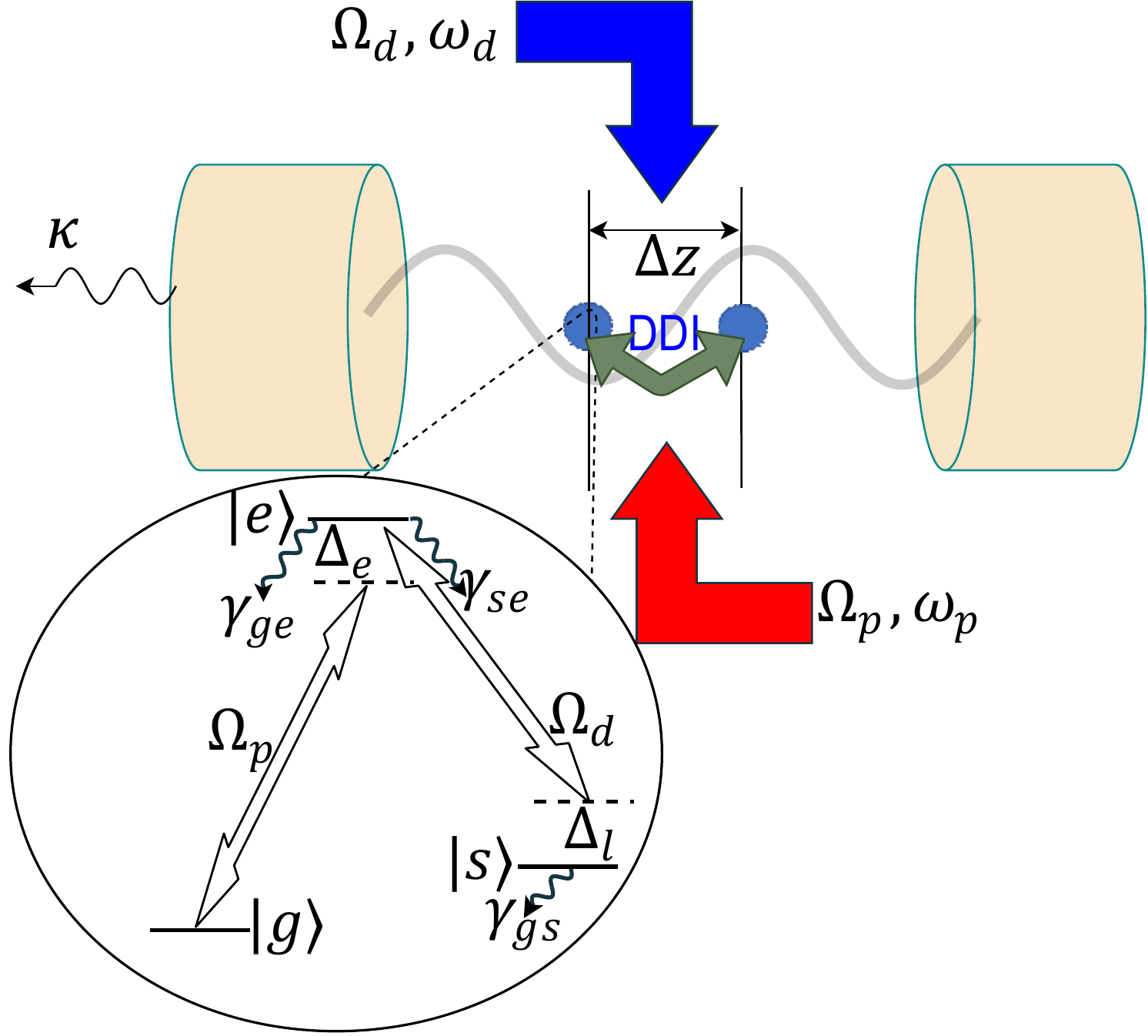}
	\caption{Schematic model of two three-level $\Lambda$-type atoms strongly coupled with a single-mode cavity of angular frequency $\omega_c$ and decay rate $\kappa$. The distance between two atoms is labeled as $\Delta z$. The spontaneous decay rates from $\ket{e}\rightarrow\ket{g}$, $\ket{e}\rightarrow\ket{s}$, and $\ket{s}\rightarrow\ket{g}$ are indicated by $\gamma_{ge}$, $\gamma_{se}$, and, $\gamma_{sg}$, respectively. }\label{fig:2atom}
\end{figure}
\section{Model and Dressed State Picture}
We consider two $\Lambda$-type three-level atoms in a single-mode optical cavity, as illustrated in Fig.~\ref{fig:2atom}.
Each atom consists of three non-degenerate energy levels $\ket{g}$, $\ket{s}$ and $\ket{e}$ [See magnified level structure in Fig.~\ref{fig:2atom}].
The transition $\ket{g}\leftrightarrow\ket{e}$ is coupled with the single mode of the cavity.
A pump field with Rabi frequency $\Omega_p$ drives this transition, whereas a drive field of Rabi frequency $\Omega_d$ is applied to the transition $\ket{e}\leftrightarrow\ket{s}$.
We also consider the  DDI between the atoms~\cite{Bargatin2000}.
The Hamiltonian of the system in a rotating frame under rotating-wave and dipole approximation is given by:
\begin{align}
H=H_0+H_I+H_d+H_P,
\end{align}
with
\begin{align}
H_0=-\hbar\sum_{i=1,2}(\Delta_e\sigma^i_{ee}+\Delta_s\sigma^i_{ss}+\Delta_ca^{\dagger}a),
\label{eq:h0}
\end{align}
\begin{align}
H_I&=\hbar[\sum_{i=1,2}g_i(a\sigma^i_{eg}+a^{\dagger}\sigma^i_{ge})+J_1(\sigma^{(1)}_{eg}\sigma^{(2)}_{ge}+H.C)\\\nonumber
&+J_2(\sigma^{(1)}_{es}\sigma^{(2)}_{se}+H.C)],
\end{align}
\begin{align}
H_d=\hbar\sum_{i=1,2}\Omega_d(\sigma^i_{es}+\sigma^i_{se}),
\label{eq:hd}
\end{align}
and
\begin{align}
H_P=\hbar\sum_{i=1,2}\Omega_p(\sigma^i_{eg}+\sigma^i_{ge}).
\label{eq:hp}
\end{align}
The bare Hamiltonian $H_0$ represents the energies of the cavity mode and atoms with detunings $\Delta_s=\omega_p-\omega_d-(\omega_s-\omega_g)$, $\Delta_l=\omega_d - (\omega_e-\omega_s)$, $\Delta_c=\omega_p-\omega_c$ and $\Delta_e=\omega_p-(\omega_e-\omega_g)=\Delta_l$+$\Delta_s$.
Here, $\omega_k$ is the frequency of state $\ket{k}(k\in [g,s,e])$ and $\omega_c$, $\omega_d$, and $\omega_p$ are the frequencies of the cavity mode, drive field, and pump field, respectively.
We assume $\omega_e-\omega_g=\omega_c$ which results in $\Delta_c=\Delta_e$.
The operator $\sigma^i_{ab}=\ket{a}^i\bra{b}(a,b \in [g,s,e])$ is used to denote the atomic transition operator for \textit{i$^{th}$} atom. 
The bosonic field annihilation (creation) operator is denoted by $a$ ($a^{\dagger}$).
The atom-field interaction and DDI between atoms is included in the Hamiltonian $H_I$ with $g_i$ the coupling strength of the \textit{i$^{th}$} atom with the field.
We consider position-dependent atom-field coupling strength $g_i=g(\cos{(2\pi z_i/\lambda_c)})$ with $z_i$ the position of $i^{th}$-atom in the cavity mode of wavelength $\lambda_c$.
The parameters $J_1$ and $J_2$ are inter-atomic dipole-dipole coupling strengths for transition $\ket{e}\leftrightarrow\ket{g}$ and $\ket{e}\leftrightarrow\ket{s}$, respectively and for mathematical simplicity it is assumed that $J_1$=$J_2=J$ and $\Delta_l$=0 such that $\Delta_e=\Delta_s=\Delta$.
The Hamiltonian $H_d$ describes the external coherent drive, whereas $H_P$ is the Hamiltonian of the pump field.
\begin{figure}[htb!]
	\includegraphics[scale=0.3]{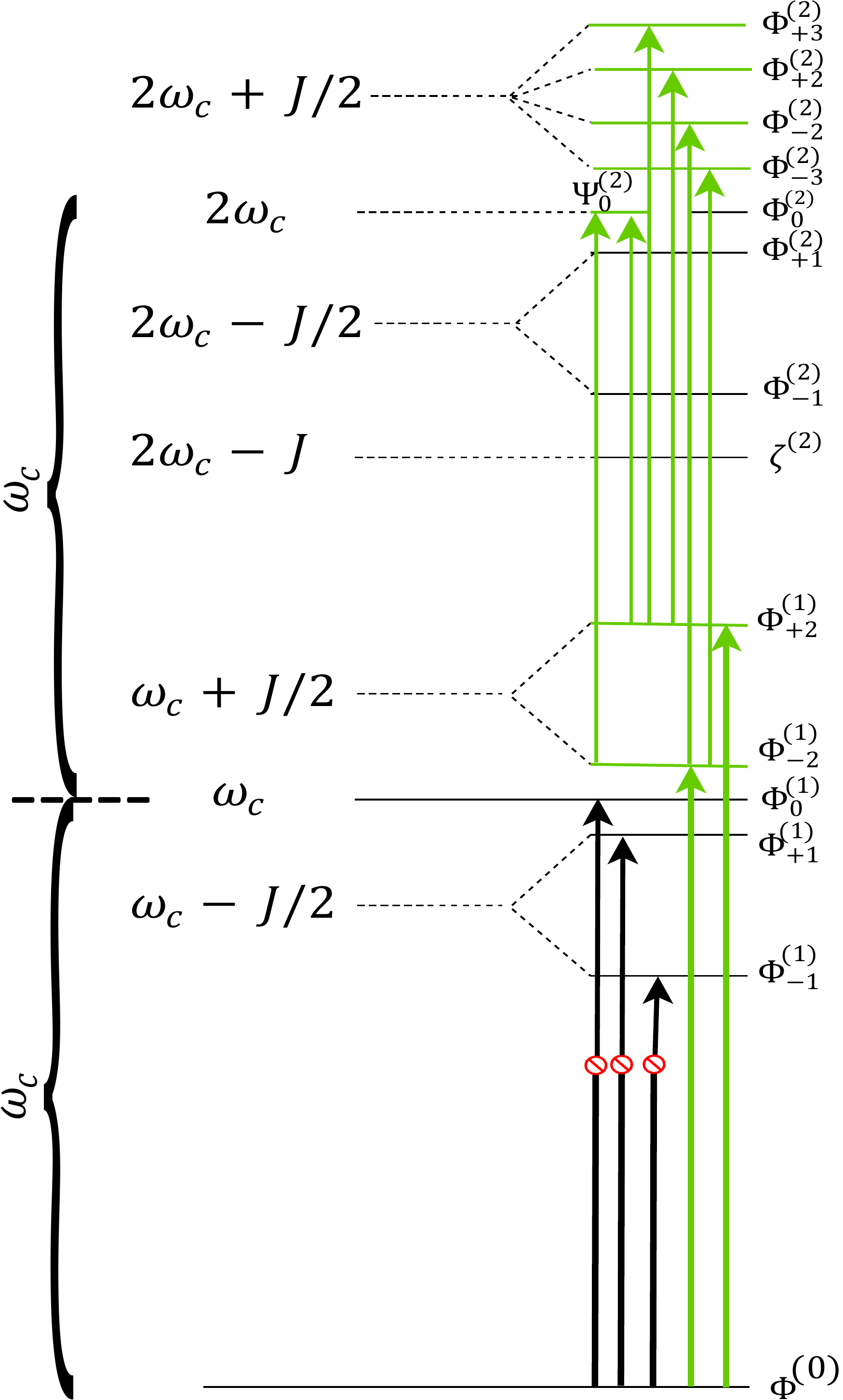}
	\caption{The dressed-state structure of important transitions on the basis of collective Dicke states for $\phi_z=0$. }\label{fig:dressed}
\end{figure}

To study the quantum properties of the system, we solve the following Lindblad master equation numerically for steady-state solutions using QuTiP~\cite{johansson2012qutip}:
\begin{equation}
\frac{d\rho}{dt}=-\frac{i}{\hbar}\left[H,\rho\right]+{\cal L}_\gamma\rho+{\cal L}_\kappa\rho,
\label{eq:1}
\end{equation}
where $\rho$ is the density matrix operator and the last two terms incorporate the atoms and cavity dissipation with rate $\gamma$ and $\kappa$, respectively.
The cavity Liouvillian function is defined as:
\begin{align}
{\cal L}_\kappa\rho=\frac{\kappa}{2}(2a\rho a^\dagger-a^\dagger a\rho-\rho a^\dagger a),
\end{align}
while atomic decay is associated with
\begin{align}
{\cal L}_{\gamma}\rho&=\frac{1}{2}\sum_{i=1,2}[\gamma_{ge}(2\sigma^i_{ge}\rho\sigma^i_{eg}-\sigma^i_{eg}\sigma^i_{ge}\rho-\rho\sigma^i_{eg}\sigma^i_{ge}) \\\nonumber
&+\gamma_{se}(2\sigma^i_{se}\rho\sigma^i_{es}-\sigma^i_{es}\sigma^i_{se}\rho-\rho\sigma^i_{es}\sigma^i_{se})\\\nonumber
&+\gamma_{gs}(2\sigma^i_{gs}\rho\sigma^i_{sg}-\sigma^i_{sg}\sigma^i_{gs}\rho-\rho\sigma^i_{sg}\sigma^i_{gs})],
\end{align}
where $\gamma_{\alpha \beta}$($\alpha,\beta$ $\in$ [\textit{e,s,g}]) is the spontaneous emission rate of state $\ket{\beta}$ to $\ket{\alpha}$.

In the absence of coherent pumping, the Hamiltonian of the system can be reformulated in collective basis states. These basis states are  $\ket{gg,1}$, $\ket{\pm^{(1)},0}$, and $\ket{\pm^{(2)},0}$ in one-photon space while in two-photon space, these are $\ket{gg,2}$, $\ket{ss,0}$,  $\ket{ee,0}$, $\ket{\pm^{(1)},1}$, $\ket{\pm^{(2)},1}$, and $\ket{\pm^{(3)},0}$ (See Appendix A for details).
The Hamiltonian matrix in one-photon space is given by
\begin{equation}
H^{(1P)}=\begin{pmatrix} \omega_c & g_+/\sqrt{2} & g_-/\sqrt{2} & 0 & 0 \\ g_+/\sqrt{2} & \omega_c +J & 0 & \Omega_d & 0\\ g_-/\sqrt{2} & 0 & \omega_c -J & 0 & \Omega_d \\ 0 & \Omega_d & 0 & \omega_c &  0 \\ 0 & 0 & \Omega_d & 0 & \omega_c \end{pmatrix}\label{eq:ops}.
\end{equation}
In two-photon space, the Hamiltonian is given by:
\begin{widetext}
\begin{align}
H^{(2P)}=\begin{pmatrix} 2\omega_c & g_+ & g_- & 0 & 0 & 0 & 0 & 0 & 0 \\ g_+ & 2 \omega_c +J & 0 & \Omega_d & 0 & 0 & 0 & 0 & g_+/\sqrt{2} \\ g_- & 0 & 2\omega_c -J & 0 & \Omega_d & 0 & 0 & 0 & -g_-/\sqrt{2} \\ 0 & \Omega_d & 0 & 2\omega_c &  0 & g_+/2 & g_-/2 & 0 & 0 \\ 0 & 0 & \Omega_d & 0 & 2\omega_c & -g_-/2 & -g_+/2 & 0 & 0\\0 & 0 & 0 & g_+/2 & -g_-/2 & 2\omega_c+J & 0 & \Omega_d & \Omega_d\\ 0 & 0 & 0 & g_-/2 & -g_+/2 & 0 & 2\omega_c-J & 0 & 0\\ 0 & 0 & 0 & 0 & 0 & \Omega_d & 0 & 2\omega_c & 0\\ 0 & g_+/\sqrt{2} & -g_-/\sqrt{2} & 0 & 0 & \Omega_d & 0 & 0 & 2\omega_c\\ \end{pmatrix}\label{eq:tps},
\end{align}
\end{widetext}
where $g_\pm=g(1\pm\cos{(\phi_z)})$ with $\phi_z$ being position-dependent phase shift between atoms and is defined as $\phi_z=2\pi\Delta z/\lambda_c$ with $\Delta z$ the distance between two atoms.
The atoms feel similar coupling with the cavity mode when $\Delta z=0$ and consequently Dicke's asymmetric states ($\ket{-^{(1)},n},\ket{-^{(2)},n},\ket{-^{(3)},n}$) become uncoupled from the cavity excitation spectrum with atoms radiating in phase. 
To characterize the single and two-photon blockades, equal time second and third order field correlation functions i.e., $g^{(2)}(0)$=$\langle a^{\dagger}a^{\dagger}a a\rangle$/$(\langle a^{\dagger} a \rangle)^2$ and $g^{(3)}(0)$=$\langle a^{\dagger}a^{\dagger} a^{\dagger} a a a\rangle$/$(\langle a^{\dagger} a \rangle)^3$ are numerically computed, respectively. 
Single-photon blockade is characterized by $g^{(2)}(0)<1$, whereas the two-photon blockade is characterized by $g^{(2)}(0)>1$ and $g^{(3)}(0)<1$.
We diagonalize Eqs.~(\ref{eq:ops}) and (\ref{eq:tps}) to obtain the energy eigenvalues and eigenstates (See Appendix B and C) to construct the dressed-state picture as shown in Fig.~(\ref{fig:dressed}). 
The criterion of allowed and forbidden transitions is associated with transition strengths by calculating the dipole matrix elements of $H_P$.

\begin{figure}[htb]
	\includegraphics[width=\linewidth]{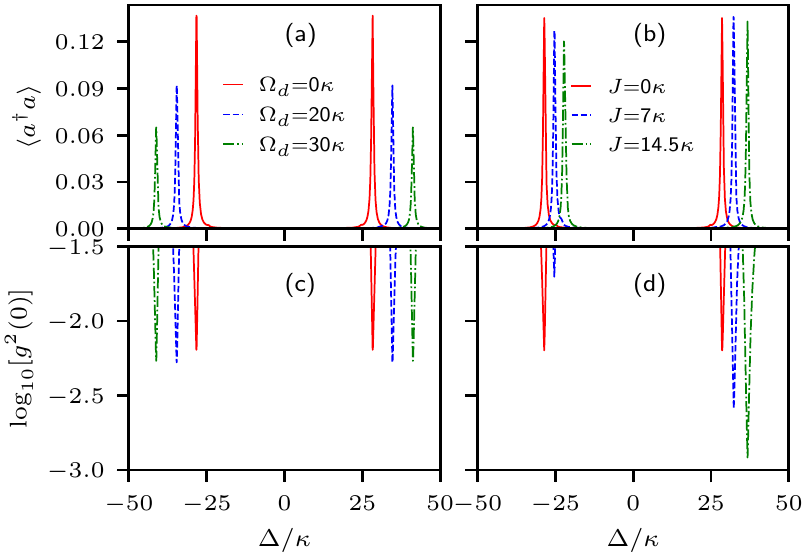}
	\caption{Mean photon number $\langle a^{\dagger}a\rangle$ and corresponding second-order field-correlation function $\log_{10}[g^{(2)}(0)]$ as a function of normalized detuning $\Delta/\kappa$ for different values of driving field strength in panel (a, c) and DDI strength in panel (b, d).  We choose $J=0$ in panel (a, c) and $\Omega_d=4\kappa$ in panel (b, d). The remaining parameters are $[\phi_z, \Omega_p, g, \gamma_{ge}=\gamma_{se},\gamma_{gs}]=[0, 0.2\kappa, 20\kappa, 0.01\kappa, \kappa]$.}\label{fig:DDI_TPB}
\end{figure}
In Fig.~(\ref{fig:dressed}), the green (black) arrows indicate the allowed (forbidden) transitions for symmetric coupling of the atoms with the cavity field ($\phi_z=0$).
The collective study of cavity-atoms shows the anharmonicity and splitting of energy levels \cite{Lin2019,Zhu2017}, however, they can be further shifted by including DDI [See Fig.~\ref{fig:dressed}]. 
Furthermore, Fig. \ref{fig:dressed} shows that for $J=0$ (no DDI), primary shifting of energy states due to DDI vanishes, and it transforms to the dressed state picture as proposed in Ref.~\cite{Han2018} for $\phi_z=0$. 
For weak coherent pumping, the system can absorb only a single photon, leading to the transitions $\Phi^{(0)}=\ket{gg,0}\rightarrow\Phi_{\pm2}^{(1)}$ (shown by green arrows in Fig.~\ref{fig:dressed}) and does not absorb the second photon as two-photon manifold is highly detuned and anharmonic.
This phenomenon is known as single-photon blockade and studied extensively recently. 
We propose in this study that allowed transitions can be shifted by including DDI as depicted in Fig.~\ref{fig:dressed} and hence affects the photon blockade strongly. 
\begin{figure}[htb!]
	\includegraphics[width=\linewidth]{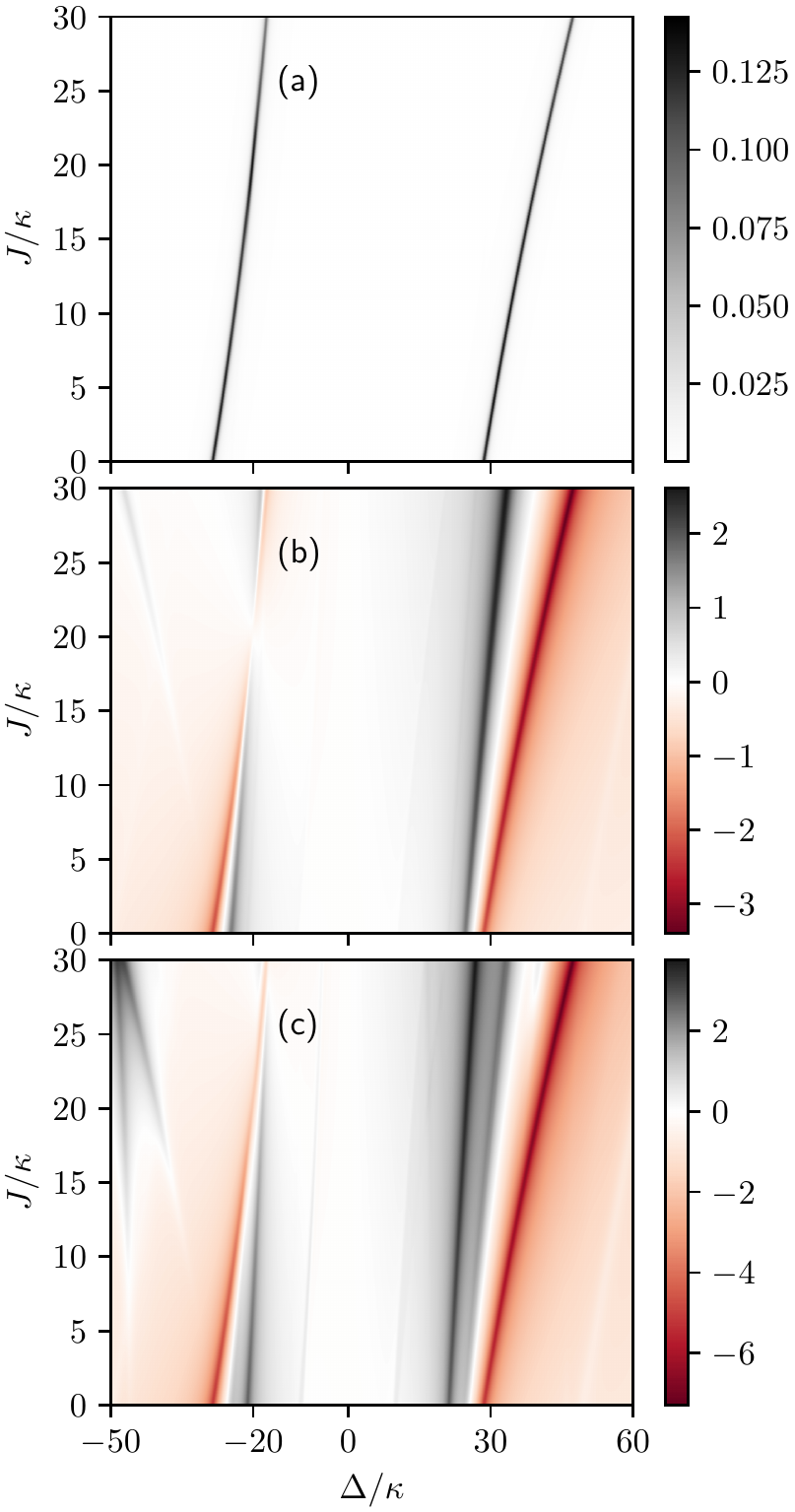}
	\caption{Mean photon number $\langle a^{\dagger}a\rangle$ (a), second-order field-correlation function $\log_{10}[g^{(2)}(0)]$ (b), and third-order field-correlation function $\log_{10}[g^{(3)}(0)]$ (c) as a function of normalized detuning $\Delta/\kappa$ and DDI strength $J/\kappa$. The rest of the parameters are the same as in Fig.~\ref{fig:DDI_TPB}(b).}\label{fig:fig4}
\end{figure}
\section{Results and Discussion}\label{sec:TPB}
In this section, we present the results of our numerical simulations.
First, we discuss the case of equal coupling of both atoms with the cavity mode (i.e., $\phi_z=0$).
We plot mean photon number ($\langle a^{\dagger}a\rangle$) and corresponding logarithmic second-order field correlation function ($g^{(2)}(0)$) as a function of normalized detuning in Fig.~\ref{fig:DDI_TPB}.
In Fig.~\ref{fig:DDI_TPB}(a), we plot the mean photon number for different values of $\Omega_d$ in the absence of DDI ($J=0$).
As the coherent pumping is weak ($\Omega_p=0.2\kappa$), one photon transitions dominate, and we obtain two symmetric peaks in $\langle a^{\dagger}a\rangle$ at frequencies $\Delta= \pm\sqrt{\Omega_d^2+2g^2}$.
This shows that the one-photon transition frequency moves away from the resonance by increasing the driving field strength ($\Omega_d$) as shown by red solid ($\Omega_d=0$), blue dashed ($\Omega_d=20\kappa$), and green dot-dashed ($\Omega_d=30\kappa$) curves in Fig.~\ref{fig:DDI_TPB}(a) (see Appendix~\ref{sec:spectrum_discussion} for further discussion).
The corresponding second-order correlation function $\log_{10}[g^{(2)}(0)]$ is plotted in Fig.~\ref{fig:DDI_TPB}(c).
It can be seen that at the frequencies $\pm\sqrt{\Omega_d^2+2g^2}$, $\log_{10}[g^{(2)}(0)]_{ \Omega_d=0\kappa}>\log_{10}[g^{(2)}(0)]_{ \Omega_d=20\kappa}\approx\log_{10}[g^{(2)}(0)]_{ \Omega_d=30\kappa}$ which shows that the strength of single-photon blockade increases by increasing driving field strength.
However, there is no significant change beyond $\Omega_d=20\kappa$.
Next, we consider the case of non-zero DDI ($J\ne0$) in Fig.~\ref{fig:DDI_TPB}(b).
We choose $\Omega_d=4\kappa$ and plot the mean photon number and second-order correlation function for different choices of $J$.
We found that the presence of DDI significantly improves the single-photon blockade.
Fig.~\ref{fig:DDI_TPB}(b) shows two asymmetric peaks in $\langle a^{\dagger}a\rangle$ at two asymmetrical frequencies $\Delta=\frac{1}{2}[J\pm\sqrt{J^2+4\Omega_d^2+8g^2}]$ whereas Fig.~\ref{fig:DDI_TPB}(d) shows the corresponding second-order correlation functions.
The strength of the DDI significantly improves the single-photon blockade at frequencies $\Delta=\frac{1}{2}[J+\sqrt{J^2+4\Omega_d^2+8g^2}]$ as $\log_{10}[g^{(2)}(0)]_{ J=0}>\log_{10}[g^{(2)}(0)]_{J=7\kappa}>\log_{10}[g^{(2)}(0)]_{J=14.5\kappa}$.
This significant improvement in single-photon blockade through DDI can only be observed in the limit $\Delta z<\lambda_c$ because within this limit DDI can be strong enough to shift and displace the energy span between one and two-photon space.

\begin{figure}[htb]
	\includegraphics[width=\linewidth]{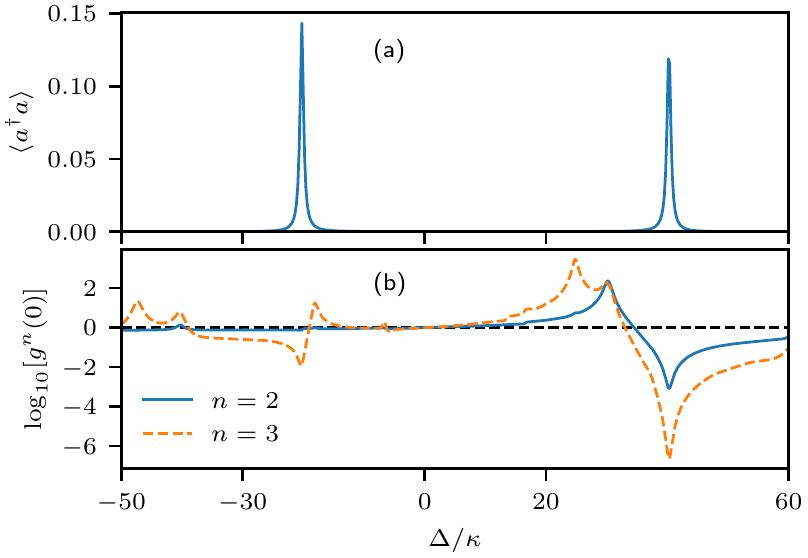}
	\caption{Mean photon number $\langle a^{\dagger}a\rangle$ (a) and corresponding second and third order field-correlation functions $\log_{10}[g^{(n)}(0)]$, $n=2, 3$ (b) as a function of normalized detuning $\Delta/\kappa$.  We choose $J/\kappa=20$ and the rest of the parameters are the same as in Fig. \ref{fig:fig4}. }\label{fig:fig5}
\end{figure}
Next, in Fig. \ref{fig:fig4}, we present a density plot of mean photon number, second-order, and third-order correlation functions against detuning and DDI strength $J$ at $\phi_z=0$.
At positive values of detuning, the second-order correlation function shows a very strong single-photon blockade with increasing DDI strength [See Fig.~\ref{fig:fig4}(b)].
However, single-photon blockade becomes very weak at negative detuning when DDI strength is increased [See Fig.~\ref{fig:fig4}(b)].
It almost vanishes when the DDI strength is comparable to the atom-field coupling strength, i.e., $J\approx g$.
This asymmetry of the spectrum and the correlation function due to the presence of DDI has an important consequence, as shown in Fig. \ref{fig:fig4}(c), where the third-order correlation function is plotted.
It can be seen that when the single-photon blockade is very weak i.e.,  $\log_{10}[g^{(2)}(0)]\approx 0$ [See Fig. \ref{fig:fig4}(b) along the mean photon spectrum peak in Fig. \ref{fig:fig4}(a)], we have a relatively weaker third-order correlation function i.e., $\log_{10}[g^{(3)}(0)] < 0$.
To illustrate better, we plot a line-cut of Fig.~\ref{fig:fig4} at $J/\kappa=20$ in Fig.~\ref{fig:fig5}.
We show the mean photon number in Fig.~\ref{fig:fig5}(a) whereas second and third-order correlation functions in Fig.~\ref{fig:fig5}(b).
At $\Delta/\kappa\approx40$, we  have strong single-photon blockade [See Fig.~\ref{fig:fig5}(b)] as discussed above.
However, the DDI-induced asymmetry in the spectrum shows that $\log_{10}[g^{(2)}(0)]\approx 0$ ($g^{(2)}(0)\approx 1$) at $\Delta/\kappa\approx-20$ showing the absence of antibunching and presence of a coherent state.
At this same value of detuning, the third-order correlation function $\log_{10}[g^{(3)}(0)] < 0$, confirming that three-photon bunching is absent.
This promises, a possibility of a two-photon bunching phenomenon if we can tune parameters to obtain $\log_{10}[g^{(2)}(0)]>0$.
Next, we show that this is indeed possible.
This is an important result that shows that by tuning the detuning of the pump field, we can achieve two very different types of photon emissions, namely single-photon emission at positive detuning and two-photon bunched emission at negative detuning.
\begin{figure}[htb]
	\includegraphics[width=0.9\linewidth]{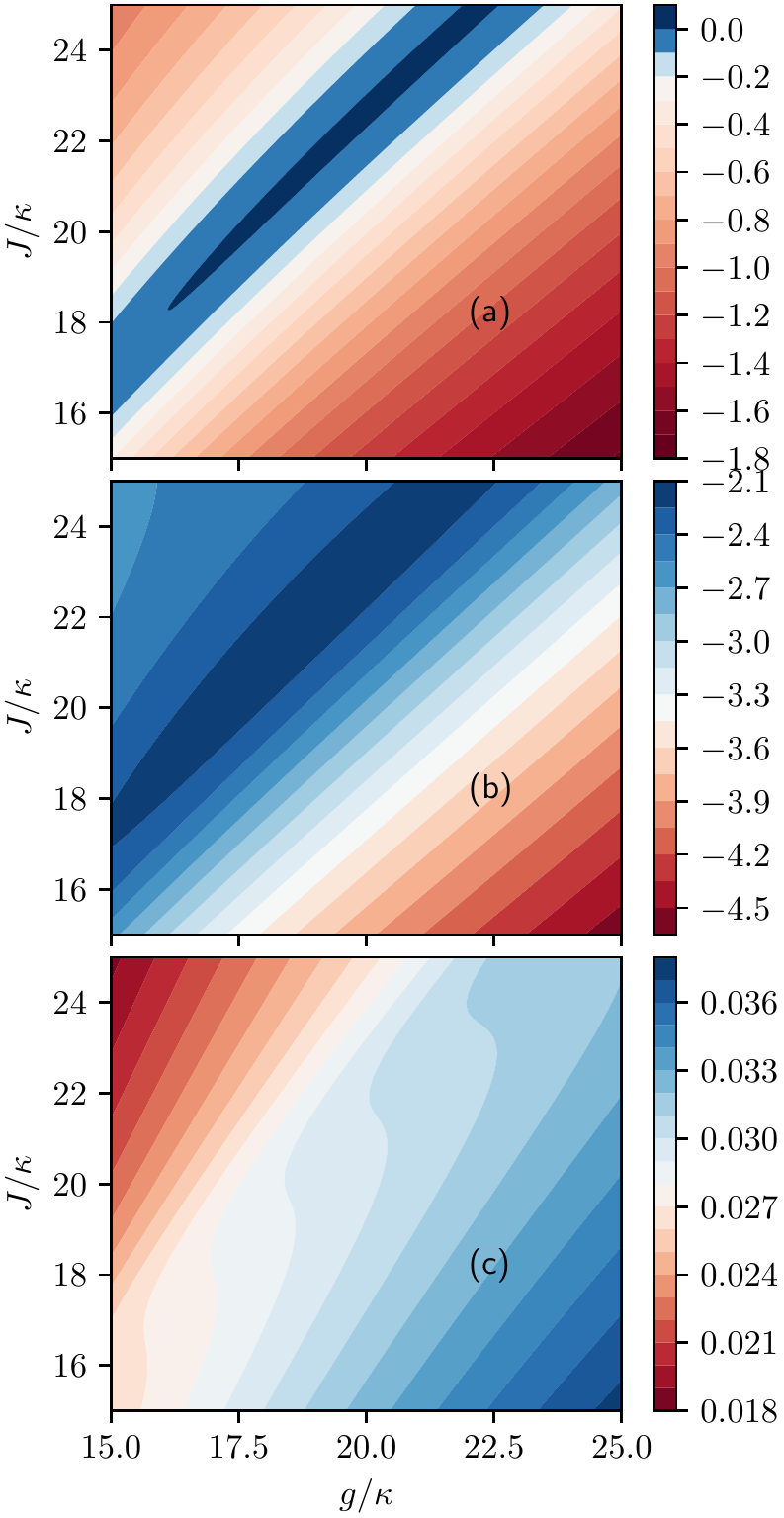}
	\caption{Second-order correlation function $\log_{10}[g^{(2)}(0)]$ (a), third-order correlation function $\log_{10}[g^{(3)}(0)]$ (b), and mean photon number $\langle a^{\dagger}a\rangle$ (c) are plotted against normalized coupling constant $g/\kappa$ and DDI strength $J/\kappa$. We choose $\Omega_d=16\kappa$ and $\Omega_p=0.1\kappa$. For each value of $g/\kappa$ and $J/\kappa$, the detuning value corresponds to the peak in the mean photon number, which is calculated using the relation given in the text. The remaining parameters are the same as in Fig. \ref{fig:fig4}.}\label{fig:fig6}
\end{figure}
\begin{figure}[htb!]
	\includegraphics[width=\linewidth]{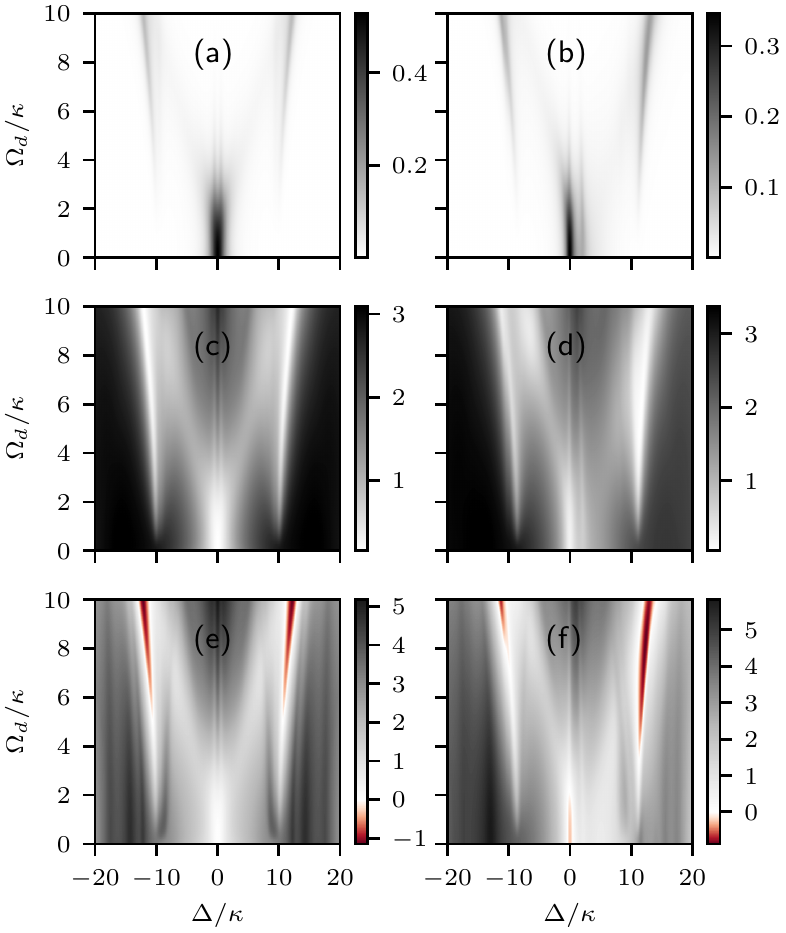}
	\caption{Mean photon number $\langle a^{\dagger}a\rangle$ (a, b), second-order correlation function $\log_{10}[g^{(2)}(0)]$ (c, d), and third-order correlation function $\log_{10}[g^{(3)}(0)]$ (e, f) as a function of $\Delta/\kappa$ and $\Omega_d/\kappa$. The system parameters are chosen as $J=0$ for left panel and $J=5\kappa$ for right panel, $\Omega_p=1.5\kappa, \phi_z=\pi$ with the remaining parameters are the same as in Fig. \ref{fig:DDI_TPB}. }\label{fig:fig7}
\end{figure}
\begin{figure}[htb!]
	\includegraphics[width=\linewidth]{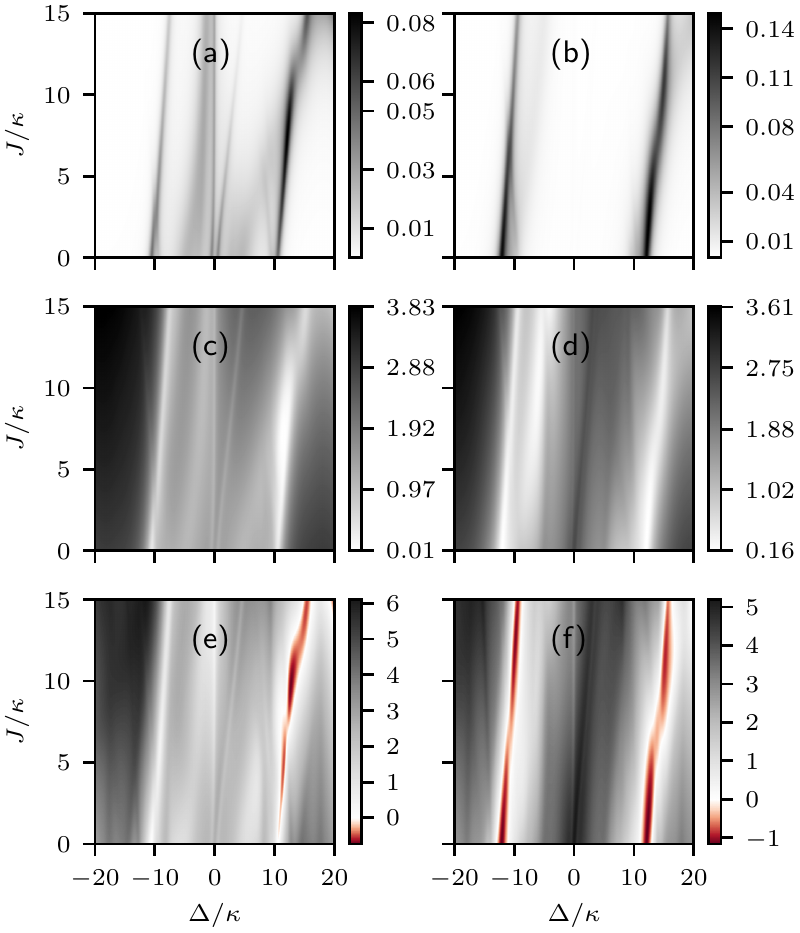}
	\caption{Mean photon number $\langle a^{\dagger}a\rangle$ (a, b), second-order correlation function $\log_{10}[g^{(2)}(0)]$ (c, d), and third-order correlation function $\log_{10}[g^{(3)}(0)]$ (e, f) as a function of $\Delta/\kappa$ and $J/\kappa$. The system parameters are chosen as $\Omega_d=5\kappa$ for the left panel and $\Omega_d=10\kappa$ for the right panel, $\Omega_p=1.5\kappa$, $\phi_z=\pi$ with the remaining parameters are the same as in Fig. \ref{fig:DDI_TPB}. }\label{fig:fig8}
\end{figure}

It is clear from Fig.~\ref{fig:fig5}(b) that the single-photon blockade almost vanishes at a particular negative detuning.
Next, we explore the parameter space where single-photon blockade completely vanishes, i.e., $\log_{10}[g^{(2)}(0)]\ge 0$.
We plot the second-order, third-order correlation functions, and mean photon number in Fig.~\ref{fig:fig6} as a function of the dimensionless coupling constant $g/\kappa$ and DDI strength $J/\kappa$.
Here, we focused on the negative detuning values, which are calculated for each combination of $g$ and $J$ using $\Delta=\frac{1}{2}[J-\sqrt{J^2+4\Omega_d^2+8g^2}]$.
We find that to obtain strictly positive values of $\log_{10}[g^{(2)}(0)]$, we need to lower the strength of $\Omega_p$.
Figure~\ref{fig:fig6}(a) shows the second-order correlation function, where the darkest region corresponds to $\log_{10}[g^{(2)}(0)]\ge 0$.
In this parameter space, single-photon blockade completely vanishes.
The slightly less dark region corresponds to $ -0.1\le \log_{10}[g^{(2)}(0)]< 0$, i.e., very weak single-photon blockade.
The corresponding third-order correlation function in Fig.~\ref{fig:fig6}(b) shows that for the region where the single-photon blockade vanishes, we have strong two-photon blockade.
It is clear that in this parameter regime, we have non-classical states of radiation field i.e., two-photon bunching with high purity, which is quite interesting.
However, reducing $\Omega_p$ also decreases the mean photon number, as shown in Fig.~\ref{fig:fig6}(c).

Finally, we consider the asymmetric coupling of the atoms with the field, i.e., $\phi_z=\pi$ and discuss the effects of DDI and drive field.
In Fig.~\ref{fig:fig7}, we plot mean photon number, second-order, and third-order correlation functions as a function of detuning and drive field strength.
In the left panel of Fig.~\ref{fig:fig7}, DDI is absent, while for the right panel, we have $J=5\kappa$.
The darker regions in Fig.~\ref{fig:fig7}(a) show the mean photon number peaks with the corresponding second and third-order correlation functions in Fig.~\ref{fig:fig7}(c) and Fig.~\ref{fig:fig7}(e), respectively.
Figure~\ref{fig:fig7}(c) shows that single-photon blockade is absent in the system for the whole range of drive field strength.
The dynamical system makes only two-photon transitions without permitting one-photon absorption as evident by $\log_{10}[g^{(2)}(0)]>0$ [See Fig.~\ref{fig:fig7}(c)] leading to the multiphoton bunching.
On the other hand, the two-photon blockade depends on the strength of the driving field and exists for $\Omega_d \gtrsim 6 \kappa$ as evident by $\log_{10}[g^{(3)}(0)]<0$ [See Fig.~\ref{fig:fig7}(e)] ensuring the absence of three-photon bunching.
Therefore, in this region of parameter space, we get two-photon bunching.
The presence of DDI [See Fig.~\ref{fig:fig7}(b, d, f)] induces an asymmetry in the spectrum showing that the range of driving field strength corresponding to two-photon blockade changes.
At positive pump field detuning, the two-photon bunching occurs for $\Omega_d \gtrsim 4 \kappa$ whereas for negative detuning it occurs for $\Omega_d \gtrsim 8 \kappa$.
The result has important consequences from an experimental point of view because the strength of the drive field required to observe two-photon bunching depends on the DDI strength.
We also note that there is also a very weak two-photon blockade (two-photon bunching) for $\Omega_d \lesssim 2 \kappa$.
Next, we show density plots of mean photon number, second-order and third-order correlation functions as a function of detuning and DDI strength in Fig.~\ref{fig:fig8}.
We choose $\Omega_d=5\kappa$ in the left panel showing that the increasing strength of DDI makes the mean photon number spectrum  more and more asymmetric around pump detuning [See Fig.~\ref{fig:fig8}(a)], while the single-photon blockade remains absent Fig.~\ref{fig:fig8}(c).
Figure~\ref{fig:fig8}(e) shows that two-photon blockade is absent at $J=0$, however, increasing DDI induces two-photon blockade at positive detuning.
Therefore, we have two-photon bunching only at positive detuning for finite DDI, which is in agreement with the behaviour shown in Fig.~\ref{fig:fig7}(e, f).
We choose a slightly stronger drive field ($\Omega_d=10\kappa$) in the right panel of Fig.~\ref{fig:fig8}.
At strong driving, we have two-photon bunching at both positive and negative pump detunings which persists for the complete range of DDI strength considered here.

Here, we briefly present the feasibility of the potential experimental realization of our proposed scheme.
The proposed scheme can be realized by placing two Rydberg atoms or ions in an optical cavity~\cite{Neuzner2016-ej, Welter_2018, Reimann2015, Vidal2007, Casabone2015, gaetan_observation_2009, urban_observation_2009, johnson_Rabi}.
Similarly, quantum dots coupled with a photonic crystal cavity or an optical microcavity is also a good candidate system for experimental realization~\cite{Kim-strong, Faraon2008, Snijders2018, Laussy_2011, jimenez-orjuela_strong_2020}.
Browaeys et al., recently reviewed the experimental realization of DDI interaction between Rydberg atoms~\cite{Browaeys_2016}.
It is shown that DDI strength $J$ in Rydberg atoms can be efficiently manipulated in experiments ~\cite{ravets_coherent_2014,barredo_2015, li_2019, shao_2017}, reaching values of the same order considered here.
Similarly, the study of exchange coupling in quantum dots is also an active area of interest~\cite{ginzel-2022,kim_exciton_2016}.
We also note that the typical values of mean photon number in microwave and optical experiments are of the order of $10^{-2}$ \cite{Hamsen2017, Snijders2018, Vaneph2018}.
We, therefore, believe, that the experimental realization of the proposed scheme is well within the reach of current experimental technology.
\section{Conclusion}
In conclusion, we studied multiphoton blockade in a single-mode cavity coupled with two three-level atoms in $\Lambda$-configuration.
We show that the presence of DDI has important consequences on the emission spectrum as well as on multiphoton blockade.
For positive values of pump field detuning for symmetric coupling of atoms, we found a positive effect of DDI, leading to a stronger single-photon blockade.
Therefore, the proposed system promises the realization of a high-purity single-photon source if a strong DDI is present.
At negative values of the pump field, DDI interaction has detrimental effects, suppressing the single-photon blockade.
We show that the single-photon blockade can be completely suppressed, accompanied by a strong two-photon blockade.
This results in the emission of non-classical photon pairs.
It is interesting to note that these two phenomena can be obtained by controlling the frequency of the pump field.
For asymmetric coupling, we show that the correct combination of drive field strength and DDI strength is important to observe two-photon bunching.
The results presented in this work are important for any potential experimental realizations where DDI between atoms is present.
%
\begin{widetext}
\appendix
\section{Definition of Collective Basis Sates}
The basis states in $n$-photon space for Eq.(\ref{eq:ops}) and Eq.(\ref{eq:tps}) are $\ket{gg,n}$, $\ket{ss,n-2}$, $\ket{ee,n-2}$,  $\ket{\pm^{(1)},n-1}$, $\ket{\pm^{(2)},n-1}$, and $\ket{\pm^{(3)},n-2}$.
The entangled states are defined as:
\begin{equation}
\ket{\pm^{(1)},n-1}=\frac{1}{\sqrt{2}}(\ket{eg,n-1}\pm\ket{ge,n-1}),
\end{equation}
\begin{equation}
\ket{\pm^{(2)},n-1}=\frac{1}{\sqrt{2}}(\ket{sg,n-1}\pm\ket{gs,n-1}),
\end{equation}
and
\begin{equation}
\ket{\pm^{(3)},n-2}=\frac{1}{\sqrt{2}}(\ket{eg,n-2}\pm\ket{ge,n-2})
\end{equation}
\section{Eigenvalues and Eigensates of Eq. (\ref{eq:ops})}
The eigenvalues and eigenstates of Eq.~(\ref{eq:ops}) for $\phi_z=0$ are given below in Table \ref{tb:tabel1}.
\begin{table}[ht]
    \centering
    \begin{tabular}{|c|c|}
    \hline
        Eigenvalues & Eigenstates\\
    \hline
       $\lambda^{(1)}_0=\omega_c$ & $\Phi^{(1)}_0=\ket{\pm^{(2)},0}-\frac{\Omega_d}{\sqrt{2}g}\ket{gg,1}$\\
    \hline
       $\lambda^{(1)}_{\pm1}=\omega_c-J/2\pm\sqrt{\frac{J^2}{4}+\Omega_d^2}$ & $\Phi^{(1)}_{\pm1}=\ket{-^{(2)},0}+\frac{\pm\sqrt{\frac{J^2}{4}+\Omega^2_d}-\frac{J}{2}}{\Omega_d}\ket{-^{(1)},0}$\\
     \hline
        $\lambda^{(1)}_{\pm2}=\omega_c+J/2\pm\sqrt{\frac{J^2}{4}+\Omega_d^2+2 g^2}$ &$\Phi^{(1)}_{\pm2}=\ket{+^{(2)},0}+\frac{\sqrt{2}g}{\Omega_d}\ket{gg,1}+\frac{1}{\Omega_d}[\frac{J}{2}\pm\sqrt{\frac{J^2}{4}+\Omega_d^2+2g^2}]\ket{+^{(1)},0}$\\
    \hline
    \end{tabular}
    \caption{One photon space}
    \label{tb:tabel1}
\end{table}
\section{Eigenvalues and Eigensates of Eq. (\ref{eq:tps})}
The two-photon manifold in Fig.~\ref{fig:dressed} is constructed based on the following eigenvalues and eigenstate of Eq.~(\ref{eq:tps}) [See Table \ref{table2}] with $A$ and $B$ defined as:
\begin{equation}
A=0.07J^2+0.43\Omega_d^2+g^2,
\end{equation}
\begin{equation}
  B=0.714 \sqrt{0.04\Omega^4_d+0.53\Omega^2_dg^2+g^4}.
  \end{equation}
\begin{table}[ht]
    \centering
    \begin{tabular}{|c|c|}
    \hline
        Eigenvalues & Eigenstates\\
    \hline
       $\lambda^{(2)}_0=\lambda^{(2)}_{0\pm}=2\omega_c$ & $\Phi^{(2)}_0,\Psi^{(2)}_0$\\
    \hline
       $\lambda^{(2)}_{\pm1}=2\omega_c-J/2\pm\sqrt{J^2/4+\Omega^2_d+g^2}$ & $\Phi^{(2)}_{\pm1}$\\
     \hline
        $\lambda^{(2)}_{\pm2}=2\omega_c+J/2\pm1.87\sqrt{A-B}$ &$\Phi^{(2)}_{\pm2}$\\
    \hline
    $\lambda^{(2)}_{\pm3}=2\omega_c+J/2\pm1.87\sqrt{A+B}$ &$\Phi^{(2)}_{\pm3}$\\
    \hline
    $\lambda^{(2)}_{\zeta}=2\omega_c-J$ & $\zeta^{(2)}$\\
    \hline
    \end{tabular}
    \caption{Two photon space}\label{table2}
\end{table}

\section{Analysis of the eigenenergy spectrum}\label{sec:spectrum_discussion}
Here, we discuss the effects of the control field and DDI on the eigenenergy spectrum. 
In the absence of the control field and DDI, we have an excitation doublet at $\Delta = \pm \sqrt{2}g$ similar to the two-level system (Ref.~\cite{Zhu2017}) corresponding to two eigenvalues of the one-photon Hamiltonian (See Table.~\ref{tb:tabel1}).
In the presence of the control field, these energy eigenstates symmetrically shift away from the resonance.
We illustrate this by plotting energy eigenvalues as a function of control field $\Omega_d$ in Fig.~\ref{fig:eigvals}.
\begin{figure}[t]
	\includegraphics[width=0.5\linewidth]{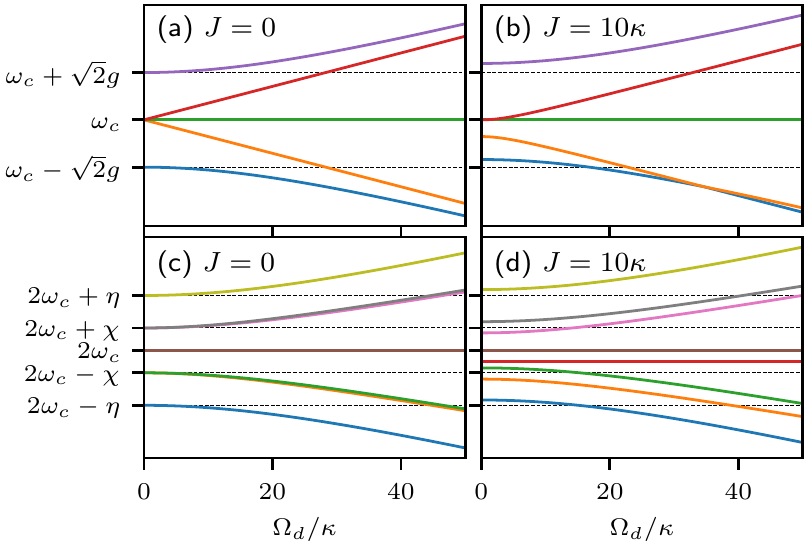}
	\caption{Energy eigenvalues of the one-photon [top panel] and two-photon [bottom panel] Hamiltonian versus Rabi frequency of the drive field. (a, c) $J=0\kappa$, (b, d) $J=10\kappa$. The parameters $\chi=1.87\sqrt{A-B}$ and $\eta=1.87\sqrt{A+B}$ are defined for $J=0$ and $\Omega_d=0$.}
 \label{fig:eigvals}
\end{figure}
In the top panel of Fig.~(\ref{fig:eigvals}), we plot eigenvalues for one-photon Hamiltonian in the absence (Fig.~\ref{fig:eigvals}(a)) and presence of DDI (Fig.~\ref{fig:eigvals}(b)).
Fig.~\ref{fig:eigvals}(a) shows that the increasing strength of the drive field symmetrically shifts the eigenenergies away from the resonance.
The presence of DDI shifts energy levels in the absence of the control field ($\Omega_d=0$), as shown in Fig.~\ref{fig:eigvals}(b).
These levels then shift asymmetrically (with respect to resonance) when the control field is applied. 
As a result, the spectrum significantly differs from Fig.~\ref{fig:eigvals}(a) where no DDI is considered.
This illustrates the control-field induced asymmetry in the spectrum at $\Delta=\frac{1}{2}[J\pm\sqrt{J^2+4\Omega_d^2+8g^2}]$ in the presence of DDI.
In Fig.~\ref{fig:eigvals}(c-d), we present the energy spectrum of the two-photon Hamiltonian (Eq.~(\ref{eq:tps})) again illustrating the anharmonicities induced by DDI.
The second asymmetry is in the strength of correlation functions at these asymmetric frequencies, leading to a single-photon blockade at positive detuning and a two-photon blockade at negative detuning.
We note that two energy eigenstates below the resonance in Fig.~\ref{fig:eigvals}(b) become almost degenerate in the presence of DDI.
This degeneracy influences the emission process and is potentially responsible for the asymmetry in correlation functions.
Furthermore, we plot, in Fig.~\ref{fig:eigvals_J}, the eigenenergy spectrum of the one-photon (Fig.~\ref{fig:eigvals_J}(a)) and two-photon (Fig.~\ref{fig:eigvals_J}(b)) Hamiltonian as a function of DDI strength $J$.
It can be seen in Fig.~\ref{fig:eigvals_J}(a) that at negative detuning, two energy levels move closer to each other as a function of $J$ and become degenerate at $J\approx g$.
Therefore, this asymmetric eigenenergy pattern in response to DDI strength is responsible for the asymmetric strengths of the correlation function.
\begin{figure}[t]
	\includegraphics[width=0.5\linewidth]{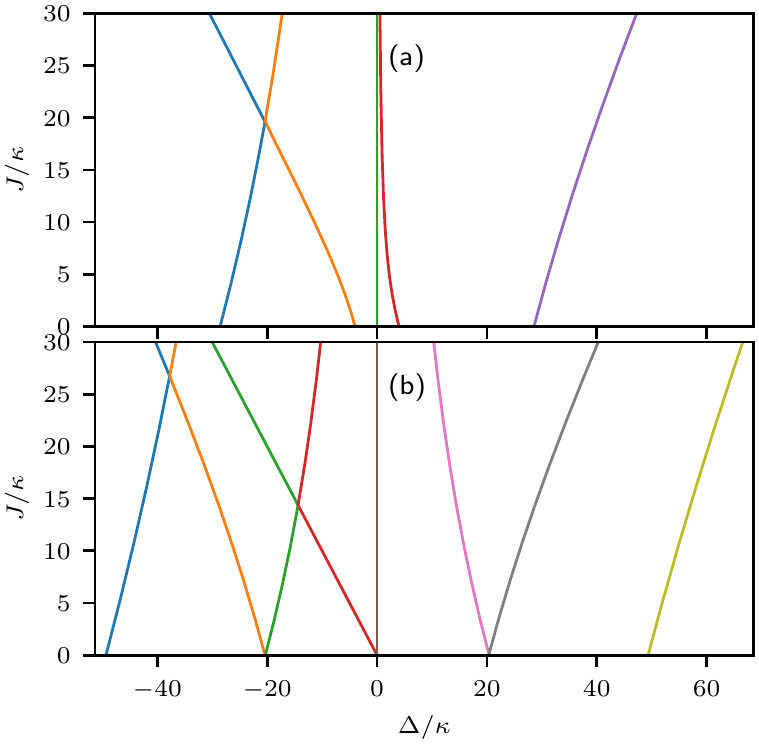}
	\caption{Energy eigenvalues of the one-photon (a) and two-photon (b) Hamiltonian versus DDI. The rest of the parameters are the same as in Fig. 4 of the main text.}\label{fig:eigvals_J}
\end{figure}

\end{widetext}
\begin{section}{Acknowledgments}
We acknowledge valuable discussions with Muhammad Waseem and Asad Mehmood.
\end{section}

\bibliography{manuscript}
\end{document}